# Features of low-to-high cycle fatigue fracture transition

**Leonid Kremnev[1], Vyacheslav Matyunin[2], Artem Marchenkov[2] and Larisa Vinogradova[3]**

[1]*MSTU «STANKIN», Russia*

[2]*NRU «MPEI», Russia*

[3] *Gubkin Russian State University of Oil and Gas (NRU), Russia*

Email: kremnevls@yandex.ru

**Abstract:** It was experimentally confirmed that if steel cyclic stress reduces to less than tensile yield stress values, i.e. in case of high-cycle fatigue, the mechanism of fracture changes from dislocation to vacancy one. This conclusion was based on the fact that steel density determined by the method of liquid displacement is less than that of steel in both initial condition and after fracture under the cyclic loads exceeding tensile yield stress values. In the latter case steel hardness increases, whereas the steels fractured under the cyclic stresses less than their tensile yield stress values show no change in hardness. It means that in such a case metal fractures without strain hardening, i.e. undergoes brittle fracturing developing by vacancy mechanism rather than by dislocation one. As a result such steel obtains the structure and properties similar to those appearing after its exposure to radiation, i.e. friability and brittleness. The results obtained allow us to explain the presence of a horizontal section in the Wöhler fatigue failure diagram.

## 1  Introduction

The hypothesis on changing the mechanism of fracture from dislocation to vacancy one under the cyclic stresses less than tensile yield strength has been experimentally confirmed by means of steel density estimation. It is demonstrated that the values of density can be used to determine and forecast the residual life period. To estimate the rate of a fatigue crack growth the author of [1] proposed to use $\nabla_{1cs}$ as the number of interatomic bonds' breakdowns at the crack tip within a single cycle of maximal variable stress $\sigma_{max}$. It was found that under $\sigma_{max} > \sigma_{0,2}$ the value of $\nabla_{1cs}$ is greater than one and amounts to several hundred of broken bonds per cycle (low-cycle fatigue); under $\sigma_{max} = \sigma_{0,2}$ we have $\nabla_{1cs} = 1$ and under $\sigma_{max} < \sigma_{0,2}$ we have $\nabla_{1cs} < 1$ (high-cycle fatigue). Since an interatomic bond can be broken only as a whole and not part by part, in the latter case it is destroyed within several (multitude of) cycles instead of a single one. In this context the mechanism of fracture may change from dislocation to vacancy one. The purpose of this work is to experimentally confirm or deny the abovementioned hypothesis. In case the density reduces in comparison with the initial value the hypothesis can be accepted; otherwise it is to be rejected.

## 2  Experimental results

To test the above assumptions a set of 3 mm thick flat "corset" specimens of 09Г2С (AISI A 516 55) and 20X13НЛ (AISI 420) steel grades were subjected to cyclic testing. Instron 8801 testing machine was used to conduct these tests at zero-to-tension stress cycle (stress ratio $R = 0$) and cycling frequency of 20 hz. A part of the specimens was tested under maximum cyclic stress $\sigma_{max}$ less than tensile yield strength, others - at $\sigma_{max}$ higher than tensile yield strength. The cyclic tests were preceded by tensile tests to determine the tensile yield strength $\sigma_T = 360$ MPa for 09Г2С and the offset tensile yield strength $\sigma_{0.2} = 255$ MPa for 20X13НЛ. After fracturing the specimens were used to cut out 35mm×30mm square blocks from their near fracture zones in order to determine their density and hardness. The density was determined by the liquid (distilled water) displacement method. The density assessment error was ± 1%. The hardness was taken as a mean value of five tests conducted on Instron Tukon 2500 machine under 1 kgf load. The error in the hardness assessment was ± 2.5%. The Table below presents the results of the experiments conducted.

Table. Cyclic fatigue tests results and fracture zone metal hardness

| Material | Maximum cycle stress $\sigma_{max}$, MPa | Up-to-fracture cycles number $N$ | Density $\rho$, g/cm$^3$ | Hardness $HV1$, kgf/mm$^2$ |
|---|---|---|---|---|
| 09Г2С | 330 ($\sigma_{max} < \sigma_{т}$) | 911×10$^3$ | 6.90 | 165 |
|  | 420 ($\sigma_{max} > \sigma_{т}$) | 60×10$^3$ | 7.40 | 213 |
|  | Metal in initial condition |  | 7.18 | 170 |
| 20Х13НЛ | 180 ($\sigma_{max} < \sigma_{0.2}$) | 98×10$^3$ | 5.70 | 188 |
|  | 300 ($\sigma_{max} > \sigma_{0.2}$) | 17×10$^3$ | 7.60 | 210 |
|  | Metal in initial condition |  | 7.27 | 190 |

The Table shows that after the cyclic tests the density of both steels has substantially changed in comparison with their initial condition. The density of 09Г2С reduced by 3,9% after testing under $\sigma_{max} < \sigma_{т}$ and increased by 3% after $\sigma_{max} > \sigma_{т}$ testing. The density of cast steel 20Х13НЛ reduced by 21.6% after testing under $\sigma_{max} < \sigma_{0.2}$ in comparison with the initial value, and increased by 4.5% after testing under $\sigma_{max} > \sigma_{0.2}$. The data presented can be referred to as an experimental validation of the hypothesis concerned. It should be noted that cast steel 20Х13НЛ shows higher spread of density values than that of wrought steel 09Г2С. This may be caused by increased amount of defects, e.g. pores and other imperfections of cast steel structure. At the same time, the values of density increase in cast and wrought steels observed after cyclic loading under maximum cycle stresses exceeding tensile yield strength are quite close to each other and as large as 3% and 4.5% respectively. When comparing the values of HV1 hardness taken before and after the cyclic tests one can see that the increase in 09Г2С hardness under $\sigma_{max} > \sigma_{т}$ was as high as 25.3%, whereas the same value for 20Х13НЛ under $\sigma_{max} > \sigma_{0.2}$ amounted to 10.5%. This increase of both steels hardness and density can be accounted for by their work-hardenability, or strain-hardening. In comparison with their initial values the changes in both steels hardness observed after their cyclic tests under $\sigma_{max}$ less than tensile yield strength appeared to be insufficient (by 1…3%) and within the limits of experimental error. This result can be deemed the second experimental evidence of the fact that strain-hardening known for its dislocation nature is absent and was replaced by vacancy mechanism of steel fracturing. It should be noted that the absence of strain-hardening signs to be observed during metallic materials fracturing speaks for its brittle nature. It should be reminded that metals and alloys are prone to increasing their vacancy amount when exposed to aggressive radioactive environments. This increase results is substantial density reduction, friability and embrittlement known as radiation friability and embrittlement. The results obtained make it possible to determine the cause for the appearance of Wöhler fatigue fracture diagram discontinuity [1]. During the vacancy fracturing a fatigue crack growth is dependent on the amount of vacancies at its tip required for this crack growth. In the period of vacancy accumulation the crack doesn't grow as the number of stress cycles increases, which results in a fatigue diagram horizontal plateau (step).

## 3    Conclusions

- In order to substantiate the hypothesis proposed it was experimentally confirmed that the mechanism of fracture changes form dislocation to vacancy one under the cyclic stresses less than tensile yield strength**,** i.e. under high-cycle fatigue conditions.
- It was found expedient to use metal density as a test parameter for metal parts and structures on-line inspection conducted to estimate their residual operation life.